\documentclass[runningheads]{llncs}

\usepackage[T1]{fontenc}

\usepackage{amssymb}

\usepackage{graphicx}

\usepackage{booktabs}

\usepackage{float}

\usepackage{url}

\usepackage{xcolor}
\usepackage{lscape}
\usepackage{caption}
\usepackage{subcaption}

\usepackage{amsmath}

\usepackage{hyperref}

\makeatletter
\g@addto@macro\normalsize{%
    \abovedisplayskip 2pt plus 1pt minus 2pt%
    \abovedisplayshortskip 2pt plus 1pt minus 2pt%
    \belowdisplayskip 2pt plus 1pt minus 2pt%
    \belowdisplayshortskip 2pt plus 1pt minus 2pt%
}
\makeatother

\title{%
  Requirements for Early Quantum Utility in the Capacitated Vehicle Routing Problem%
}
\titlerunning{Requirements for Quantum Advantage in CVRP}

\author{%
  Chinonso Onah\inst{1,2}\orcidID{0000-0002-6296-533X}%
  \and Kristel Michielsen\inst{3, 4}\orcidID{0000-0003-1444-4262}%
}

\authorrunning{C. Onah and K. Michielsen}

\institute{%
  Volkswagen AG, Berliner Ring 2, Wolfsburg 38440, Germany \\%
  \email{chinonso.calistus.onah@volkswagen.de} \\%
  \and Department of Physics, RWTH Aachen University, Aachen, Germany \\%
  \and J\"ulich Supercomputing Centre, Forschungszentrum J\"ulich, Germany \\
  \and Universit\"at zu K\"oln, 50923 K\"oln, Germany
}

\begin{document}
\maketitle
\begin{abstract}
We introduce a transparent, encoding-agnostic framework for determining when the Capacitated Vehicle Routing Problem (CVRP) can achieve \emph{early quantum advantage}. Our analysis shows that this is unlikely on noisy-intermediate-scale quantum (NISQ) hardware even in the best case scenario utilizing the most efficient encoding models.  Closed-form resource counts combined with the latest device benchmarks yield three decisive “go/no-go’’ figures of merit—the quantum feasibility point plus the qubit- and gate-feasibility lines—that place any CVRP instance on a single decision diagram.  Contrasting a direct QUBO mapping with the space-efficient higher-order (HOBO) encoding reveals a stark gap. Applied to early-advantage benchmarks such as \texttt{$Golden_5$}, our diagram shows that HOBO circuits require merely $7685$ whereas their QUBO counterparts still exceed 200\,000 qubits.  In addition to identifyiing probable candidate instances for Early Quantum Advantage in CVRP, our framework therefore provides the first unifying “go/no-go’’ metric that ingests any CVRP encoding alongside any hardware profile and highlights precisely when quantum devices could challenge classical heuristics.
\end{abstract}



\section{Introduction}\label{sec:intro}
The Capacitated Vehicle Routing Problem (CVRP) is a cornerstone of logistics:  
for a fleet of even a few hundred vehicles, a 1–2\% cost improvement translates into seven-figure annual savings and thousands of tonnes of \(\mathrm{CO_2}\) emissions \cite{zeyen2025shiftingburdensdelayeddecarbonisation}.  
Classical exact methods stall beyond \(n\!\approx\!120\) customers \cite{baldacci2023stateoftheart}, while large real-world instances still carry double-digit optimality gaps after hours of CPU time (Table~\ref{tab:cvrp_gaps}).  Quantum heuristics promise better scaling, yet two obstacles persist:
(i)~direct QUBO mappings demand \(k(n+1)^2\) qubits—orders of magnitude beyond today’s hardware, and  
(ii)~the community lacks a unified yard-stick to declare an instance ``feasible'' on a given device.  We close this gap by:
\begin{enumerate}
  \item deriving closed-form qubit and depth counts for both naïve QUBO and space-efficient HOBO encodings;
  \item extracting hardware-level randomisation thresholds \((N_{\max},D_{\max})\) from recent benchmarking data~\cite{MontanezBarrera2024b}; and
  \item combining (1) and (2) into a single feasibility diagram that answers, at a glance, whether a given CVRP instance and encoding can run meaningfully on present-day or near-term quantum processors.
\end{enumerate}

The rest of paper is organised as follows: Section \ref{sec:intro} formally states the CVRP and  defines the qubit count of both encodings.  
Section \ref{sec:feas} defines the feasibility metrics and plots them against benchmark instances.  
Section \ref{sec:high} quantifies the business value of small gap reductions while Section \ref{sec:res} presents the resource estimation for CVRP and identifies some candidate problems for early quantum advantage. In the appendix \ref{sec:small_feas} we provide insight into small scale problems suitable for algorithmic development and Section \ref{sec:conclusion} concludes.

\subsection{Problem Statement}
\label{sec:intro}
To formulate the Capacitated Vehicle Routing Problem, consider a weighted graph \(G = (V, E)\) where:
\begin{itemize}
    \item \(V = \{0,1,2,\ldots,n\}\) is the set of nodes with \(0\) denoting the depot, and nodes \(1,2,\ldots,n\) are customers.
    \item Each customer node \(i \in \{1,\ldots,n\}\) has demand \(q_i\).
    \item there are \(k\) vehicles, each with capacity \(C\).
    \item the weight of the edge $(i,j)$, \(w_{ij}\) denotes the cost of traveling from node \(i\) to node \(j\). 
\end{itemize}
The Capacitated Vehicle Routing Problem\cite{laporte2009fifty} (CVRP) asks for a set of routes \(\{r_1, r_2, \ldots, r_k\}\), one route per vehicle, such that:
\begin{enumerate}
    \item Each customer is visited exactly once
    \item For each vehicle \(m\), the sum of demands of the customers on its route does not exceed \(C\).
    \item Each route starts and ends at the depot (node \(0\)).
    \item The total cost \(\sum_{m=1}^{k} \sum_{(i,j)\in r_m} w_{ij}\) is minimized.
\end{enumerate}
\label{sec:proposed_approach}

For every ordered pair $(i,j)\in V\times V$ and vehicle $v\in\{1,\dots,k\}$ we introduce  
\[
  x_{ij}^v=\begin{cases}
    1 & \text{if vehicle $v$ traverses edge $(i,j)$,}\\
    0 & \text{otherwise}.
  \end{cases}
\]

Mathematically, one writes:
\begin{align}
\text{minimize} \quad & \sum_{v=1}^{k} \sum_{i,j \in V} w_{ij} \, x_{ij}^v, \label{eq:cvrp-cost} \\
\text{subject to} \quad & \sum_{v=1}^{k} \sum_{j \in V} x_{ij}^v = 1, \quad \forall i \in V \setminus \{0\}, \label{eq:visit-once}\\
& \sum_{j \in V} x_{0j}^v = 1, \quad \forall v, \label{eq:depart-depot}\\
& \sum_{i \in V} x_{i0}^v = 1, \quad \forall v, \label{eq:return-depot}\\
& \sum_{i,j \in V} q_i \, x_{ij}^v \le C, \quad \forall v, \label{eq:capacity}
\end{align}

In the direct QUBO Encoding of CVRP\cite{Lucas2014,Feld2019,Palackal2023}, a direct mapping of each $x_{ij}^v$ to a qubit leads to $Q_{\text{q}} = k \times (n)^2$ qubits.
When capacity constraints are fully encoded the QUBO approach requires an additional $k \cdot C$ qubits. We keep the depot as vertex $0$, therefore the total number of vertices is $n+1$, elevating the total qubit count to:
\begin{equation}
Q_{\mathrm{q}}=k[(n+1)^2+C]
\label{eq:naive-qubits}
\end{equation}

In the Space-Efficient, Higher Order Binary Optimization (HOBO) Encoding\cite{glos2020spaceefficient,bentley2022quantum}, a more compact representation uses binary expansions for vehicle indices and for capacity encoding. For example, to encode route permutations, we use $\log(n)$ qubits per node (per vehicle). In practice one employs the \emph{permutation-gadget} of
Glos\,\cite{glos2020spaceefficient} layered with the position-register trick of
Bentley\,\cite{bentley2022quantum}: each city is assigned
$\lceil\log_2 n\rceil$ ancilla qubits that store its position in the tour, so the register Hilbert space grows merely as $n\,2^{\lceil\log_2 n\rceil}$
instead of $n!$. Similarly, capacity can be monitored by $\log(C+1)$ qubits rather than $C$ qubits. This leads to the qubit count:
\begin{equation}
Q_{\text{h}} = k \left[ (n)\log(n) \right] + k \left[\log(C+1)\right].
\label{eq:hobo-qubits}
\end{equation}
Although this significantly reduces qubit counts, higher-order polynomials emerge in the cost function, demanding multi-qubit gates and higher circuit depths in quantum circuits.

\begin{figure}[t]
  \centering
  \begin{subfigure}[b]{0.48\linewidth}
    \centering
    \includegraphics[width=\linewidth]{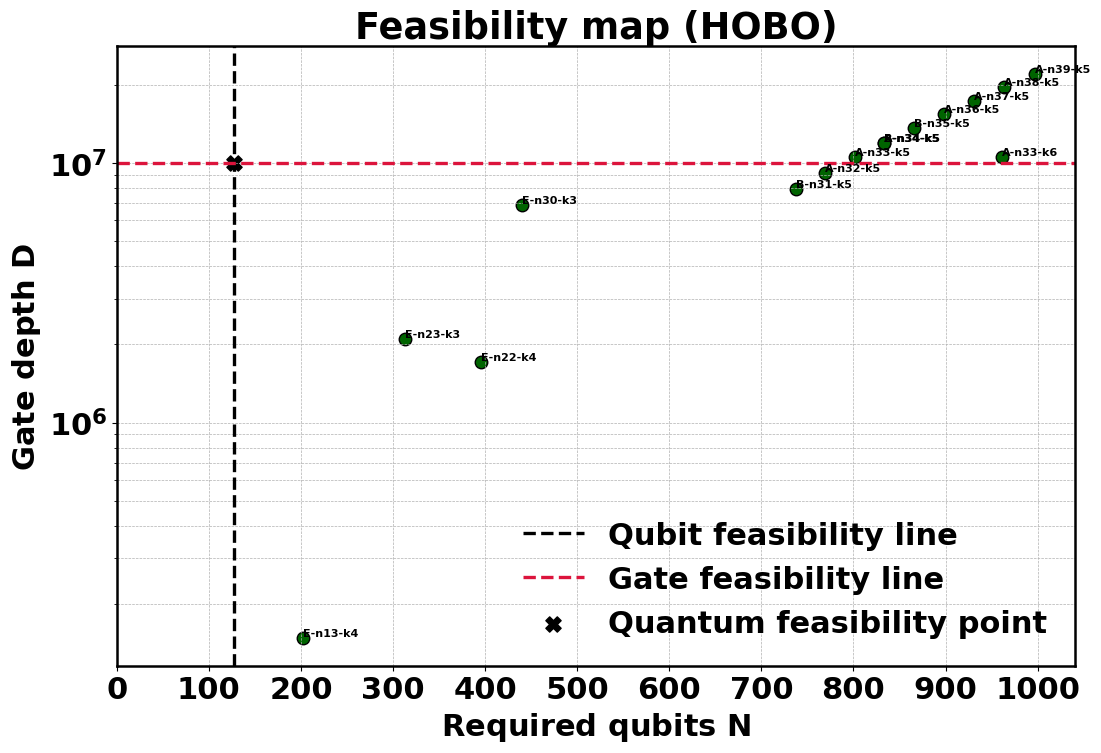}
    \caption{CVRPLib instances}
    \label{fig:feas-prototypes}
  \end{subfigure}
  \hfill
  \begin{subfigure}[b]{0.48\linewidth}
    \centering
    \includegraphics[width=\linewidth]{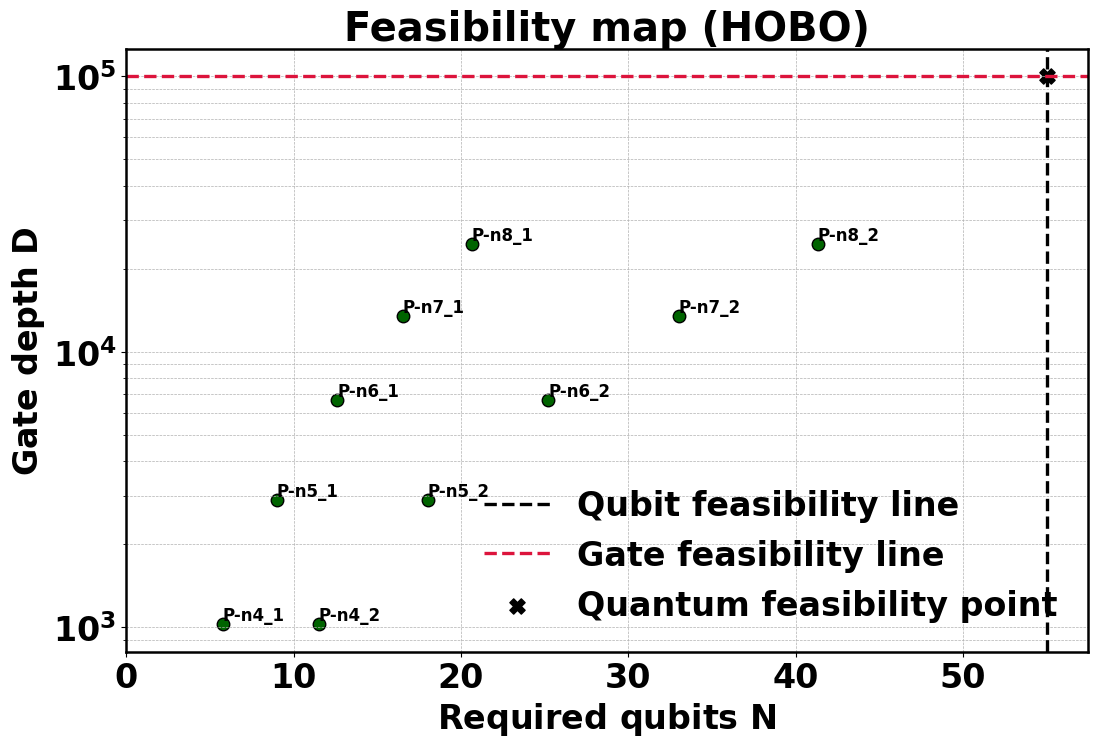 }
    \caption{QOPTLib instances}
    \label{fig:feas-qoptlib}
  \end{subfigure}

  \caption{\textbf{Feasibility maps for HOBO-encoded CVRP instances.}
  Each green dot represents a Capacitated Vehicle Routing Problem (CVRP) instance placed by its required logical-qubit count $N$ (horizontal axis) and two-qubit-gate depth $D$ (vertical axis) when encoded with the higher-order binary optimisation (HOBO) Hamiltonian.  
  The \emph{vertical dashed line} marks the projected qubit budget $N_{\max}$ for next-generation gate-based devices 
  The \emph{horizontal dashed line} shows a conservative gate-depth ceiling $D_{\max}=10^{7}$, inferred from recent QPU benchmarking studies.  
  The black star identifies the \emph{quantum-feasibility point}, i.e.\ the hardware operating point at $(N_{\max},D_{\max})$.  
  Instances that fall \emph{below and to the left} of both lines are executable on the assumed hardware; those above or to the right require further maturation but can be tracked along the feasibility lines as quantum technology improves.  
  Together, the two panels reveal that although no current benchmark instance is yet feasible, many lie only one to two hardware generations away, making them useful yardsticks for monitoring quantum-hardware progress.}
  \label{fig:combined_fes}
\end{figure}

\subsection{Quantum Feasibility}
\label{sec:feas}

A fundamental observation in current NISQ devices is that beyond certain limits of qubit count ($N$) and circuit depth ($D$), hardware outputs start to resemble a uniformly random distribution~\cite{MontanezBarrera2024b}. For the sake of hardware probes, we capture this phenomenon via a \textbf{threshold function} $f(N, D)$. Formally, define $f(N, D) = 0$ if the device outputs maintain a fidelity above some cutoff $\tau$, and $f(N, D) = 1$ otherwise. Recent benchmarking studies by Montañez-Barrera \emph{et al.} reported in Ref.~\cite{MontanezBarrera2024b} demonstrate that every contemporary quantum device has an inherent \emph{randomization threshold}. Specifically, when the qubit number \( N \) and the gate depth \( D \) exceed this threshold, the output state degenerates into a random distribution. Consequently, meaningful computation becomes impossible, as the resulting measured bitstrings no longer encode interpretable solutions.

We introduce the quantum feasibility metrics as follows:

\begin{definition}[Quantum Feasibility Point]
\label{def: def1}
The \textbf{Quantum Feasibility Point} \((N_{\max}, D_{\max})\) is defined as the maximum allowable combination of qubit number \( N \) and circuit depth \( D \) for which a quantum device can reliably perform meaningful computations. Here, the \(N\)-axis represents the number of qubits, and the \(D\)-axis denotes the number of two-qubit gates utilized in the quantum circuit.
\end{definition}

\begin{definition}[Qubit Feasibility Line]
\label{def: def2}
The \textbf{Qubit Feasibility Line} \( N_{\max} \) represents the maximum number of qubits a quantum device can handle before crossing the randomization threshold, given a fixed gate depth \(D\).
\end{definition}

\begin{definition}[Gate Feasibility Line]
\label{def: def3}
The \textbf{Gate Feasibility Line} \( D_{\max} \) indicates the maximum gate depth achievable by a quantum device before approaching  randomization, for a fixed number of qubits \(N\).
\end{definition}

The intersection of these two feasibility lines, \( (N_{\max}, D_{\max}) \), precisely defines the quantum feasibility point, marking the boundary within which reliable quantum computations are feasible.

The feasibility point $(N_{\max},D_{\max})$ is a \emph{hardware-level} criterion:  
if an \emph{ideal} circuit requires more than $N_{\max}$ qubits \emph{or} more than $D_{\max}$ native two-qubit gates, the physical device decoheres into essentially uniform noise before \emph{any} algorithmic information can be imprinted on the wave-function.  
Whether the algorithm \emph{itself} attains a low optimality gap is a \emph{separate} (higher-level) question that can then be studied independently.

If a CVRP instance's associated circuit (under a given encoding) requires more than \((N_{\max},D_{\max})\), that instance is classified as \emph{infeasible} for that hardware, regardless of the theoretical polynomial runtime of the quantum algorithm. We also note that the feasibility picture sketched in this way is conservative because it ignores post-processing techniques like quantum error mitigation that effectively extend $D_{\max}$. In Fig.~\ref{fig:combined_fes} the dashed lines could potentially shift when these techniques are implemented.

\subsection{Motivation}

Exploration of quantum computing for combinatorial optimization is advancing rapidly, with new heuristics and exact approaches emerging almost daily. Yet, in the subdomain of vehicle routing and mobility optimization, published results remain relatively scarce. This is especially true for the Capacitated Vehicle Routing Problem (CVRP), a fundamental testbed in transportation and logistics optimization, where existing quantum-solution studies have been confined to highly trivial instances (e.g., four or five customer locations served by a single vehicle) \cite{gambella2020quantum}.

The core challenge arises from the dominant practice of using QUBO encodings, wherein classical binary decision variables are directly mapped to qubits. Although this is polynomially efficient from a complexity-theoretic standpoint, it is suboptimal in practice. For example, even the smallest instance in the well-known CVRPLIB demands a minimum of 5305 qubits—effectively rendering such prototypes infeasible on current and near-term quantum hardware.

Fortunately, emerging approaches in space-efficient problem encoding \cite{glos2020spaceefficient} offer a promising alternative. By employing binary encodings for permutations and capacities, the qubit requirement can be dramatically reduced, transforming CVRP benchmark instances—once out of reach —into candidates for implementation on current or near-term devices. In particular, we will see that this transformation brings all Vehicle routing sub problems listed in QOPTLib\cite{Osaba_2023} within the feasibility range for current devices without circuit cutting or problem decomposition.
 
\paragraph{The feasibility metric} is intended for  
(i) \textbf{algorithm designers}, who can favour encodings that plot \emph{below} the line;  
(ii) \textbf{hardware teams}, who aim to push the line \emph{upward}; and  
(iii) \textbf{logistics practitioners}, who can time their adoption based on the
intersection of their problem sizes with these shifting boundaries.

\section{CVRP as a High-Value Application}
\label{sec:high}
High-value applications of quantum computing are those where even incremental improvements in solution quality can yield substantial aggregated economic, environmental, or operational benefits. Formally, we define a high-value quantum computing application as follows:

\begin{definition}[High-Value Quantum Computing Applications]
\label{def:high}
A quantum computing application is considered \textbf{high-value} if a modest enhancement \( \Delta \) in the quality of solutions (e.g., optimality gap reduction, energy consumption, or resource utilization) yields significant aggregated impacts in terms of cost reduction, operational efficiency, and environmental sustainability.
\end{definition}

A prominent candidate among such high-value problems is the Capacitated Vehicle Routing Problem (CVRP) because in practical scenarios, slight reductions in CVRP costs have significant financial cost and environmental implications. For instance, in a large fleet with total annual distances of around 100 thousand kilometers, a modest improvement of 2\% corresponds to an annual saving of thousands of kilometers; directly saving thousands of dollars in fuels and vehicle maintainance costs and thousands of tonnes of CO\(_2\) emissions; potentially contributing to the emission reduction challenges in the transport sector. \cite{zeyen2025shiftingburdensdelayeddecarbonisation}. For a parcel-delivery fleet of $2\,500$ vans (typical of European operators) a 2 \% mileage reduction corresponds to  
$\approx\!1.8$ million km, \$1.2 M in diesel costs, and \( \sim4,600 \,\mathrm{t}\,\mathrm{CO_{2}e}\) annually\footnote{Assumes 30\,L/100 km and 2.6 kg CO\(_2\)/L.}.  
Even a \(\Delta\mathrm{gap}=1\%\) improvement on instances listed in Table~\ref{tab:cvrp_gaps} thus yields seven-figure savings, satisfying our Definition ~\ref{def:high}.

\begin{figure}[ht]
    \centering
    \begin{subfigure}[t]{0.49\columnwidth}
        \centering
        \includegraphics[width=\textwidth]{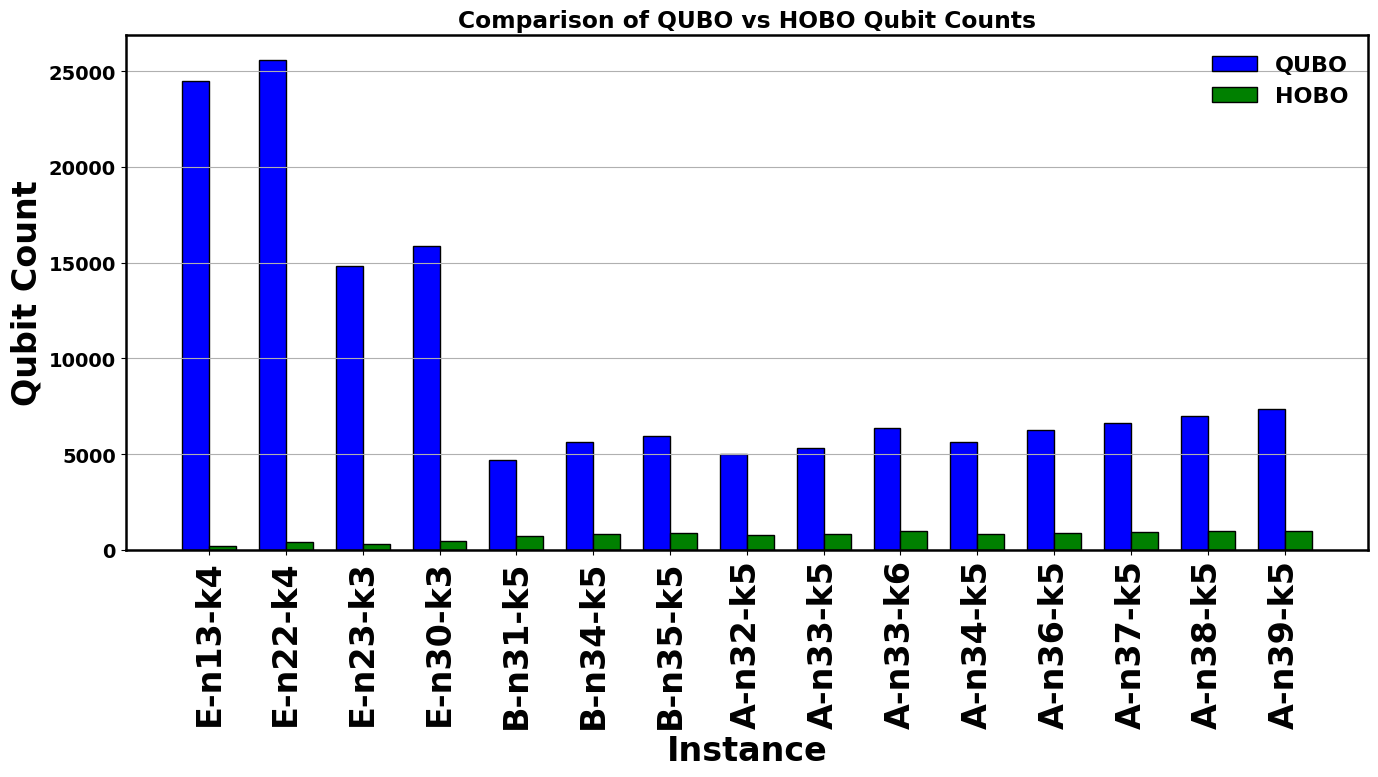}
        \caption{Select Instances from CVRPLib }
        \label{fig:qubo_cost}
    \end{subfigure}%
    \hfill
    \begin{subfigure}[t]{0.48\columnwidth}
        \centering
        \includegraphics[width=\textwidth]{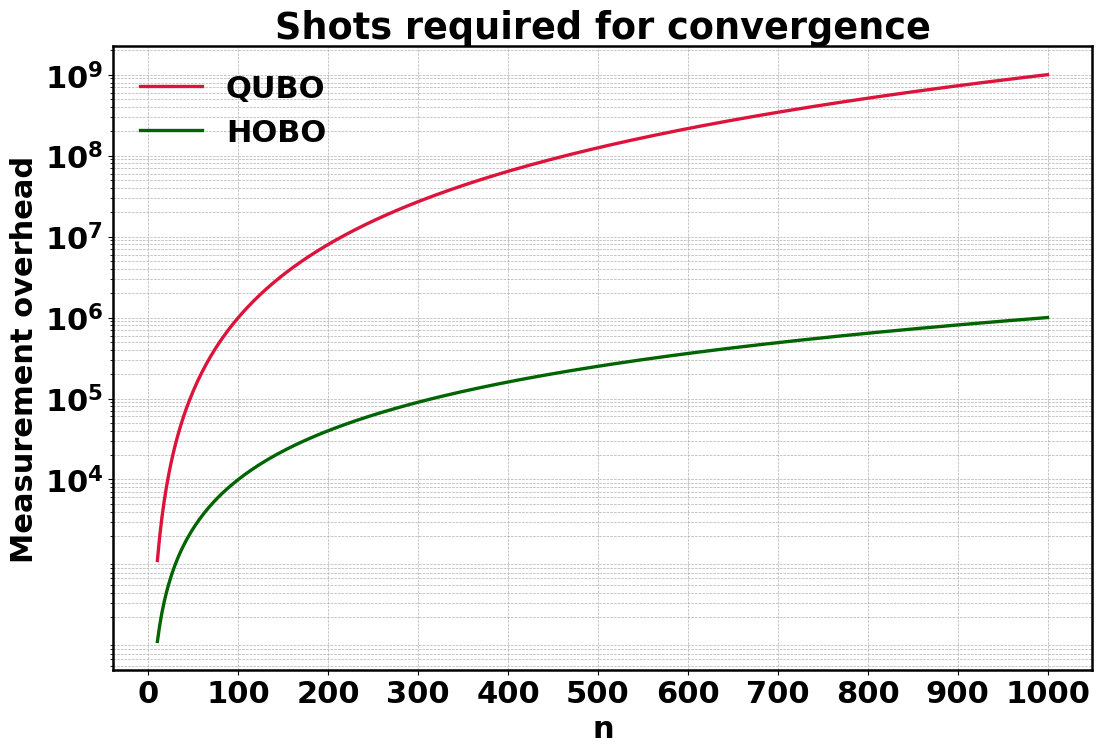} 
        \caption{Measurement Overhead}
        \label{fig:ratio_decay}
    \end{subfigure}
    \caption{Quantum‐resource requirements and feasibility outlook for solving prototype instances of the Capacitated Vehicle Routing Problem (CVRP) on gate‐based quantum hardware. Table~\ref{tab:hamiltonian-encodings} compares two Hamiltonian encodings of CVRP—quadratic unconstrained binary optimisation (QUBO) and higher‐order binary optimisation (HOBO).  Relative to the canonical QUBO mapping, HOBO compresses the qubit register from $N^{2}$ to $N\log N$ and modestly increases the number of Hamiltonian terms from $\mathcal{O}(N^{3})$ to $\tfrac12 N^{4}$.  
  These savings propagate to circuit volume (product of circuit depth and number of qubits) and to the number of distinct Pauli measurements (b), which drop from $\mathcal{O}(N^{3})$ to $\mathcal{O}(N^{2})$.  
  Circuit depth $D$ scales polynomially with the instance size $N$ via $D(\cdot)\!\propto\!\operatorname{poly}(N)$, so the principal trade‐off of the HOBO encoding is a higher (but still polynomial) gate complexity.  These problems still remain out of reach of near term devices as can be seen from Figure~\ref{fig:combined_fes}
  }
    \label{fig:combined}
\end{figure}

\section{Quantum Resource Estimation for CVRP}
\label{sec:res}

The pursuit of quantum advantage in combinatorial optimization, particularly within the Capacitated Vehicle Routing Problem (CVRP), necessitates accurate estimation of quantum computational resources. In Eq.~\ref{eq:naive-qubits} and \ref{eq:hobo-qubits}, we presented estimates of the qubit cost for solving CVRP via QUBO and HOBO methods, respectively. To complete the quantum resource estimation and determine quantum feasibility, we further require an approximate assessment of the number of gates necessary.

\begin{table}[ht]
\centering
\begin{tabular}{lcc}
\hline
 & ${\text{QUBO}}$ & ${\text{HOBO}}$ \\
\hline
No.\ of qubits       & $N^2$                & $N \log(N)$ \\
No.\ of terms        & $2N^3$               & $\tfrac{1}{2} N^4$ \\
Circuit volume       & $12N^3$              & $2N^4 \log(N)$ \\
No.\ of measurements & $\mathcal{O}(N^3)\,\max_{i,j} W_{i,j}$ 
                     & $\mathcal{O}(N^2)\,\max_{i,j} W_{i,j}$ \\
\hline
\end{tabular}%
\caption{Asymptotic estimates of resources required for various Hamiltonian encodings from Ref. ~\cite{glos2020spaceefficient}.}
\label{tab:hamiltonian-encodings}
\end{table}

Quantum circuit depth \(D\), typically scales polynomially with the instance size \(N\) and and is directly related to the number of terms in the Hamiltonian\cite{glos2020spaceefficient}:
\begin{equation}
D(.) \propto \text{poly}(N).
\label{eq:circuit_depth}
\end{equation}
The relevant expressions are summarized in Table~\ref{tab:hamiltonian-encodings}. Clearly, the number of terms in the Hamiltonian, quantum volume and circuit depths scale differently for QUBO and HOBO encodings. Circuit complexity is the price paid for the exponential savings with the HOBO method. 

However, despite leading to a higher circuit complexity, the HOBO method not only significantly reduces the required number of qubits (from \(N^2\) to \(N \log(N)\)), but also substantially decreases the number of measurements required—from \(\mathcal{O}(N^3)\) in QUBO to \(\mathcal{O}(N^2)\) in HOBO. These advantages make HOBO particularly attractive, despite its increased and complexity. It is also worth noting that on gate-based quantum computers, custom proprietry multi-qubit interactions are reportedly being developed to implement these multi-qubit HOBO interactions directly\cite{bentley2022quantum}  thereby circumventing extra gate-depth complexity. This opens the gate to leveraging the exponential advantage in encoding illustrated in Fig.~\ref{fig:combined}.


\section{Conclusion}
\label{sec:conclusion}

We have introduced a systematic method for evaluating quantum feasibility in solving Capacitated Vehicle Routing Problems (CVRP). By formulating explicit quantum feasibility metrics based on qubit count and circuit depth, we offer practical guidelines to assess hardware readiness for tackling specific CVRP instances. Our analysis demonstrates how advanced encodings, such as HOBO, substantially lower qubit requirements compared to naive encodings, bringing previously infeasible problem instances within reach of contemporary quantum processors despite the increase in circuit depth. Future improvements in gate fidelity is expected to expand these feasibility boundaries. Although quantum advantage in industrial-scale routing problems remains aspirational, our feasibility framework clearly delineates the boundaries of current quantum capabilities, providing industry practitioners a transparent roadmap for quantum hardware maturity. Our contribution  translates \emph{any} chosen encoding and the latest hardware data into a single, actionable decision boundary. As quantum hardware continues to improve, ongoing monitoring of these feasibility metrics will enable stakeholders to make informed, strategic decisions regarding investments in quantum computing technology and its integration into operational workflows.

\section*{Data Availability} All data in this paper are derived within the paper.

\section*{Conflict of Interests}
All authors declare no competing interests.

\section*{Author contributions}
C.O. wrote the main paper and prepared all the figures, K.M. provided guidance during the project. All the authors reviewed the paper.

\section*{Funding}
This research received no external funding.

\section*{Additional information}
All our results are reported in the main paper.

\textbf{Correspondence and requests for materials} should be addressed to C.O.\ (\texttt{chinonso.calistus.onah@volkswagen.de}).


\newpage
\appendix
\section{\textbf{Appendix}}

\subsection{Opportunities for Quantum Utility in CVRP}
The \emph{high-value} nature of CVRP indicates an immense potential for quantum utility but what does it take to realize this? It is already known that exact methods struggle for instances approaching $100$ customer locations and classical heuristics methods remain unable to solve instances with fewer than $1000$ customer locations to provable optimality. Recent computational surveys show that branch-and-cut methods close the gap only for $n\!\lesssim\!120$ and capacitated instances with $n\!>\!500$ still have certified gaps $\!>\!10\%$ after days of CPU time~\cite{baldacci2023stateoftheart,toth2024vrpsurvey}.  To our knowledge \emph{no} instance with more than $1\,000$ customers and capacities has been solved to proven optimality.

The performance of classical heuristics and approximation methods is typically quantified through the \textit{optimality gap}, defined as:

\begin{equation}
\text{Gap}(\%) = 100 \times \frac{\text{solution value} - \text{lower bound}}{\text{lower bound}}
\end{equation}
Where the lower bounds can be calculated from the LP relaxation that can be iteratively tightened to estimate how far away a given solution is from optimality. For small problems, this iterative tightening from below (through the LP relaxation) and above (through the heusristic solution) eventually leads to $0\%$ gaps for problems solved to optimality. In the classical solution certification method above, eventually either the $lower-bound$ or the $solution-value$ stops improving and the solution method is stuck. In Table~\ref{tab:cvrp_gaps} we indicate the currently \emph{Best Known Solutions} (BKS) for moderately sized problems and the possible gaps to optimality computed from $LP$ relaxations of the problems.

These significant gaps provide a compelling rationale for exploring quantum solutions, where even incremental reductions in optimality gaps could substantially enhance operational efficiency, cost-effectiveness, and environmental sustainability. 

The exploration of quantum frameworks, such as Quantum Annealing\cite{Feld2019}, the Quantum Approximate Optimization Algorithm (QAOA)\cite{Farhi2014QAOA,xie2024feasibilitypreserved,Palackal2023,bentley2022quantum}, Quantum Phase Estimation Algorithm\cite{NielsenChuang2000} and Quantum Search via Grover Algorithm\cite{Grover1996}, for addressing CVRP is thus motivated not only by theoretical interest but also by the significant real-world benefits achievable in high-value applications.

\subsection{VRP Instances for Algorithmic Development}
\label{sec:small_feas}
It is however crucial that algorithmic development for CVRP continues in anticipation of quantum hardware maturity. To this end, small Vehicle Routing Problem instances were introduced in QOPTLib\cite{Osaba_2023}. While QUBO method would render most of them infeasible on current hardwares, the space efficient encoding renders some of them feasible on the state of the art Superconducting and Trapped Ion quantum hardwares as indicated in Figure \ref{fig:combined_fes}.

Thus, the feasibility limits described above could serve as a concise \textbf{hardware maturity metric} for industrial stakeholders. It could serve as \emph{Progress Indicator} because as improvements in gate fidelity or partial error correction push these feasibility limits and monitoring these changes helps in mapping hardware progress to specific application prototypes. Firms evaluating quantum solutions for large-scale CVRP can decide whether to proceed with current devices or to wait for next-generation QPUs that extend the operational regime.

Although these prototype CVRP instances are classically solvable, the corresponding quantum circuits (even under a space-efficient encoding) are not efficiently classically simulable due to the exponential scaling of the quantum state space. This dichotomy makes them an excellent dataset for prototyping quantum solutions: the exact classical solutions can be used to benchmark the performance of quantum algorithms, while the quantum circuits themselves, being too complex for efficient classical simulation, serve as a challenging testbed for hardware performance and error mitigation techniques.

\subsection{Current Feasibility Estimates for CVRP}
\label{sec:feasibility_lines}
Extending the asymptotic analyses in Table~\ref{tab:hamiltonian-encodings} to  moderately sized CVRP instances whose optimal solutions remain unknown allows us to identify what could be regarded as CVRP candidates for \emph{\textbf{early quantum advantage}}. Table~\ref{tab:challenging_cvrp} presents a quantum resource estimation results for these special CVRP instances. Some of them are as small as 200 customer locations and yet, their optimal solutions remain unknown for decades, setting them apart as true candidates to test new algorithms and technologies that promise some advantage compared to established classical procedures.

For the standard CVRP prototypes, one can easily estimate their feasibility in a straightforward way with Equations \ref{eq:naive-qubits}, and \ref{eq:hobo-qubits} and data from Table~\ref{tab:hamiltonian-encodings}. In the recent QPU benchmarking studies~\cite{MontanezBarrera2024b} the maximum number of two qubit gates before randomization occured on the best quantum processing unit was found to be in the range $10^7$. Thus in accordance with Definitions \ref{def: def1}, \ref{def: def2},\ref{def: def3} we set   $D_{max}= 10^7$ to estimate where quantum feasibility points lie depending on whether the number of qubits needed would fit on the hardware and if the number of terms in the problem formulation translates to number of gates less than $D_{max}$.

Because of the quadratic dependence on $n$ and linear dependence on $C$, even the smallest problems in the CVRPLIB, Capacitated Vehicle Routing Problem Library of benchmark instances needs at least $5305$ qubits to encode in the QUBO method. This is significantly greater than $N_{max}$. The only alternative is the HOBO approach but these problems are still not currently feasible because they require $200$ to $1000$ qubits in addition to number of gates exceeding $D_{max}$. But one could hope that they are within the reach of next generation of quantum devices with $400$  to $1200$ qubits. 

A plot of the Qubit feasibility lines and Gate feasibility line for these prototype instances are presented Fig. \ref{fig:combined_fes}. While these instances are currently not in the feasible range, their feasibility lines  can be tracked over time by industry practiioners to monitor hardware maturity.

\section{Quantum Resource Estimation Tables}
\begin{table}[ht!]
\centering
\caption{Optimality Gaps in Selected CVRP Instances}
\label{tab:cvrp_gaps}        
\setlength{\tabcolsep}{4pt} 
\begin{tabular}{lccc}
\toprule
\textbf{Instance}   & \textbf{BKS} & \textbf{Lower Bound} & \textbf{Gap (\%)} \\
\midrule
Loggi-n401-k23      & 336\,903  & 261\,353.7  & 22.42 \\
Loggi-n501-k24      & 177\,078  & 101\,513.3  & 42.67 \\
Loggi-n601-k19      & 113\,155  &  75\,710.84 & 33.09 \\
Loggi-n601-k42      & 347\,059  & 201\,159.5  & 42.04 \\
Loggi-n901-k42      & 246\,301  & 132\,381.5  & 46.25 \\
Loggi-n1001-k31     & 284\,356  & 158\,041.4  & 44.42 \\
ORTEC-n242-k12      & 123\,750  & 101\,608.6  & 17.89 \\
ORTEC-n323-k21      & 214\,071  & 168\,276.5  & 21.39 \\
ORTEC-n405-k18      & 200\,986  & 159\,327.7  & 20.73 \\
ORTEC-n455-k41      & 292\,485  & 217\,255.7  & 25.72 \\
ORTEC-n510-k23      & 184\,529  & 153\,962.2  & 16.56 \\
ORTEC-n701-k64      & 445\,543  & 339\,019.9  & 23.91 \\
\bottomrule
\end{tabular}
\end{table}

\begin{table}[ht!]
\centering
\caption{Challenging CVRP Instances. Qubit counts and circuit depths rely on asymptotic assumptions.   Treating the numbers in Table~\ref{tab:hamiltonian-encodings} as the \emph{per‐layer} cost of a parity‐check mixer within the Quantum Approximate Optimisation Algorithm (QAOA), and assuming a constant or slowly growing layer count and  
extending the asymptotic formulas to moderate‐scale instances whose optimal solutions are still unknown ($\sim200$ customers) identifies a regime where CVRP may deliver \emph{early quantum advantage}.
  }
\label{tab:challenging_cvrp}
\small
\resizebox{\columnwidth}{!}{%
\begin{tabular}{lrrrrrrrr}
\toprule
\textbf{Problem Instance} & 
\textbf{n} & 
\textbf{Vehicles} & 
\textbf{Cap.} & 
\textbf{QUBO} & 
\textbf{HOBO} & 
\textbf{Depth(N)} & 
\textbf{Quantum Vol.} & 
\textbf{Error Rate} \\
\midrule
\texttt{Loggi-n401-k23}   & 400 & 23 & 100 & 3663923 & 79649 & 398245 & 31719816005 & 3.1e-11 \\
\texttt{ORTEC-n242-k12}   & 241 & 12 & 125 & 692700  & 22956 & 114780 & 2634889680  & 3.8e-10 \\
\texttt{ORTEC-n323-k21}   & 322 & 21 & 100 & 2165961 & 56448 & 282240 & 15931883520 & 6.3e-11 \\
\texttt{ORTEC-n405-k18}   & 404 & 18 & 160 & 2926242 & 63072 & 315360 & 19890385920 & 5.0e-11 \\
\texttt{X-n280-k17}       & 279 & 17 & 192 & 1317092 & 38641 & 193205 & 7465634405  & 1.3e-10 \\
\texttt{X-n303-k21}       & 302 & 21 & 794 & 1919295 & 52416 & 262080 & 13737185280 & 7.3e-11 \\
\texttt{X-n308-k13}       & 307 & 13 & 246 & 1220466 & 33059 & 165295 & 5464487405  & 1.8e-10 \\
\texttt{X-n327-k20}       & 326 & 20 & 128 & 2115060 & 54560 & 272800 & 14883968000 & 6.7e-11 \\
\texttt{X-n359-k29}       & 358 & 29 & 68  & 3697993 & 88247 & 441235 & 38937665045 & 2.6e-11 \\
\texttt{X-n411-k19}       & 410 & 19 & 216 & 3182443 & 67735 & 338675 & 22940151125 & 4.4e-11 \\
\texttt{Golden\_1}         & 240 & 9  & 550 & 519039  & 17154 & 85770  & 1471298580  & 6.8e-10 \\
\texttt{Golden\_2}         & 320 & 10 & 700 & 1024610 & 26720 & 133600 & 3569792000  & 2.8e-10 \\
\texttt{Golden\_3}         & 400 & 9  & 900 & 1440909 & 31194 & 155970 & 4865328180  & 2.1e-10 \\
\texttt{Golden\_4}         & 480 & 10 &1000 & 2304410 & 42840 & 214200 & 9176328000  & 1.0e-10 \\
\texttt{Golden\_5}         & 200 & 5  & 900 & 202505  & 7685  & 38425  & 295296125   & 3.3e-09 \\
\texttt{Golden\_6}         & 280 & 7  & 900 & 551187  & 15995 & 79975  & 1279200125  & 7.8e-10 \\
\texttt{Golden\_7}         & 360 & 8  & 900 & 1038248 & 24528 & 122640 & 3008113920  & 3.3e-10 \\
\texttt{Golden\_8}         & 440 & 10 & 900 & 1936210 & 38720 & 193600 & 7496192000  & 1.3e-10 \\
\texttt{Golden\_9}         & 255 & 14 &1000 & 917224  & 28658 & 143290 & 4106404820  & 2.4e-10 \\
\texttt{Golden\_10}        & 323 & 16 &1000 & 1674944 & 43216 & 216080 & 9338113280  & 1.1e-10 \\
\texttt{Golden\_11}        & 399 & 17 &1000 & 2709868 & 58752 & 293760 & 17258987520 & 5.8e-11 \\
\texttt{Golden\_12}        & 483 & 19 &1000 & 4433156 & 81985 & 409925 & 33607701125 & 2.9e-11 \\
\texttt{Golden\_17}        & 240 & 22 & 200 & 1261062 & 41888 & 209440 & 8773022720  & 1.1e-10 \\
\bottomrule
\end{tabular}%
}
\end{table}


\begin{thebibliography}{99}


\bibitem{zeyen2025shiftingburdensdelayeddecarbonisation}
E.~Zeyen, S.~Kalweit, M.~Victoria, and T.~Brown,
``Shifting burdens: how delayed decarbonisation of road transport affects other sectoral emission reductions,''
\emph{Environmental Research Letters}, vol.~20, no.~4, article~044044, 2025.
doi: \url{https://doi.org/10.1088/1748-9326/adc290}.


\bibitem{baldacci2023stateoftheart}
R.~Baldacci, P.~Toth, and D.~Vigo,
``Exact algorithms for routing problems under vehicle capacity constraints,''
\emph{Annals of Operations Research}, vol.~175, no.~1, pp.~213--245, 2010.
doi: \url{https://doi.org/10.1007/s10479-009-0650-0}.



\bibitem{MontanezBarrera2024b}
J.~A. Monta\~nez-Barrera, K.~Michielsen, and D.~E.~Bernal-Neira,
\newblock ``Evaluating the performance of quantum process units at large width and depth,''
\newblock \emph{arXiv preprint arXiv:2502.06471}, 2025.
\newblock doi: \url{https://doi.org/10.48550/arXiv.2502.06471}.


\bibitem{laporte2009fifty}
G.~Laporte,
\newblock ``Fifty years of vehicle routing,''
\newblock \emph{Transportation Science}, vol.~43, no.~4, pp.~408--416, 2009.
\newblock doi: \url{https://doi.org/10.1287/trsc.1090.0301}.


\bibitem{Farhi2014QAOA}
E.~Farhi, J.~Goldstone, and S.~Gutmann,
\newblock ``A Quantum Approximate Optimization Algorithm,''
\newblock \emph{arXiv preprint arXiv:1411.4028}, 2014.
\newblock \url{https://arxiv.org/abs/1411.4028}.

\bibitem{Lucas2014}
A.~Lucas,
``Ising formulations of many NP problems,''
\emph{Frontiers in Physics}, vol.~2, article~5, 2014.
doi: \url{https://doi.org/10.3389/fphy.2014.00005}.

\bibitem{Feld2019}
S.~Feld, C.~Roch, T.~Gabor, C.~Seidel, F.~Neukart, I.~Galter, W.~Mauerer, and C.~Linnhoff-Popien,
``A Hybrid Solution Method for the Capacitated Vehicle Routing Problem Using a Quantum Annealer,''
\emph{Frontiers in ICT}, vol.~6, article~13, 2019.
doi: \url{https://doi.org/10.3389/fict.2019.00013}.

\bibitem{Palackal2023}
L.~Palackal, B.~Poggel, M.~Wulff, H.~Ehm, J.~M.~Lorenz, and C.~B.~Mendl,
``Quantum-Assisted Solution Paths for the Capacitated Vehicle Routing Problem,''
in \emph{Proc. of the 2023 IEEE International Conference on Quantum Computing and Engineering (QCE)}, 
vol.~1, IEEE, 2023, pp.~648--658.
doi: \url{https://doi.org/10.1109/QCE57702.2023.00080}.
\newblock \url{https://arxiv.org/abs/2304.09629}.




\bibitem{glos2020spaceefficient}
A. Glos, A. Krawiec, and Z. Zimborás,
``Space-efficient binary optimization for variational quantum computing,''
\emph{npj Quantum Information} \textbf{8}, 39 (2022).
doi: \url{https://doi.org/10.1038/s41534-022-00546-y}.


\bibitem{xie2024feasibilitypreserved}
N.~Xie, X.~Lee, D.~Cai, Y.~Saito, N.~Asai, and H.~C.~Lau,
``A feasibility-preserved quantum approximate solver for the Capacitated Vehicle Routing Problem,''
\emph{Quantum Information Processing}, vol.~23, no.~8, article~291, 2024.
doi: \url{https://doi.org/10.1007/s11128-024-04497-5}.


\bibitem{gambella2020quantum}
S.~Harwood, C.~Gambella, D.~Trenev, A.~Simonetto, D.~Bernal~Neira, and D.~Greenberg,
``Formulating and solving routing problems on quantum computers,''
\emph{IEEE Transactions on Quantum Engineering}, vol.~2, article~3100118, pp.~1--17, 2021.
doi: \url{https://doi.org/10.1109/TQE.2021.3049230}.




\bibitem{NielsenChuang2000}
M.~A. Nielsen and I.~L. Chuang,
``Quantum Computation and Quantum Information,''
Cambridge, UK: Cambridge University Press, 2000.


\bibitem{Grover1996}
L.~K. Grover,
``A fast quantum mechanical algorithm for database search,''
in \emph{Proceedings of the 28th Annual ACM Symposium on Theory of Computing (STOC)},
ACM, 1996, pp.~212--219.
doi: \url{https://doi.org/10.1145/237814.237866}.



\bibitem{toth2024vrpsurvey}
K.~Braekers, K.~Ramaekers, and I.~Van~Nieuwenhuyse,
``The vehicle routing problem: State of the art classification and review,''
\emph{Computers \& Industrial Engineering}, vol.~99, pp.~300--313, 2016.
doi: \url{https://doi.org/10.1016/j.cie.2015.12.007}.



\bibitem{bentley2022quantum}
C. D. B. Bentley, S. Marsh, A. R. R. Carvalho, P. Kilby, and M. J. Biercuk,
``Quantum computing for transport optimization,''
arXiv:2206.07313 [quant-ph] (2022).
\url{https://arxiv.org/abs/2206.07313}


%
\bibitem{Willsch_2020} 
M. Willsch, D. Willsch, F. Jin, H. De Raedt, and K. Michielsen, 
``Benchmarking the quantum approximate optimization algorithm,'' 
\emph{Quantum Information Processing} \textbf{19}, no. 7 (2020). 
DOI: \href{http://dx.doi.org/10.1007/s11128-020-02692-8}{10.1007/s11128-020-02692-8}.

\bibitem{Michielsen_2017} 
K. Michielsen, M. Nocon, D. Willsch, F. Jin, T. Lippert, and H. De Raedt, 
``Benchmarking gate-based quantum computers,'' 
\emph{Computer Physics Communications} \textbf{220}, 44--55 (2017). 
DOI: \href{http://dx.doi.org/10.1016/j.cpc.2017.06.011}{10.1016/j.cpc.2017.06.011}.


\bibitem{TemmeZNE}
K.~Temme, S.~Bravyi, and J.~M. Gambetta,
\newblock ``Error Mitigation for Short-Depth Quantum Circuits,''
\newblock \emph{Physical Review Letters}, vol. 119, 180509, 2017.

\bibitem{Osaba_2023}
E.~Osaba and E.~Villar-Rodriguez,
\newblock ``QOPTLib: A Quantum Computing Oriented Benchmark for Combinatorial Optimization Problems,''
\newblock in \emph{Benchmarks and Hybrid Algorithms in Optimization and Applications}, Springer Nature Singapore, 2023, pp. 49–63.

\bibitem{osaba2024hybridquantumsolversproduction}
E.~Osaba, E.~Villar-Rodriguez, A.~Gomez-Tejedor, and I.~Oregi,
\newblock ``Hybrid Quantum Solvers in Production: how to succeed in the NISQ era?,''
\newblock \emph{arXiv preprint arXiv:2401.10302}, 2024.
\bibitem{Wallman2016}
J.~J. Wallman and J.~Emerson,
``Noise tailoring for scalable quantum computation via randomized compiling,''
\emph{Phys. Rev. A}, 94:052325, 2016.







\end{thebibliography}
\end{document}